# An integrated assessment of the impact of precipitation and groundwater on vegetation growth in arid and semiarid areas

Lin Zhu [1 2], Huili Gong [1], Zhenxue Dai[2], Tingbao Xu [3], Xiaosi Su[4]

[1] *College of Resources Environment and Tourism, Capital Normal University, Laboratory Cultivation Base of Environment Process and Digital Simulation, Beijing 100048, China*

[2] *Earth and Environmental Sciences Division, Los Alamos National Laboratory, Los Alamos, New Mexico 87545, United States*

[3] *Fenner School of Environment and Society, Australian National University, Canberra ACT 0200, Australia*

[4] *College of Environment and Resources, Jilin University, Changchun 130021, China*

**Abstract:** Increased demand for water resources together with the influence of climate change has degraded water conditions which support vegetation in many parts of the world, especially in arid and semiarid areas. This study develops an integrated framework to assess the impact of precipitation and groundwater on vegetation growth in the Xiliao River Plain of northern China. The integrated framework systematically combines remote sensing technology with water flow modeling in the vadose zone and field data analysis. The vegetation growth is quantitatively evaluated with the remote sensing data by the Normalized Difference Vegetation Index (NDVI) and the simulated plant water uptake rates. The correlations among precipitation, groundwater depth and NDVI are investigated by using Pearson correlation equations. The results provide insights for understanding interactions between precipitation and groundwater and their contributions to vegetation growth. Strong correlations between groundwater depth, plant water uptake and NDVI are found in parts of the study area during a ten-year drought period. The numerical modeling results indicate that there is an increased correlation between the groundwater depth and vegetation growth and that groundwater significantly contributes to sustaining effective soil moisture for vegetation growth during the long drought period. Therefore, a decreasing groundwater table might pose a great threat to the survival of vegetation during a long drought period.

*index; numerical simulation; plant water uptake; northern China*

## 1. Introduction

Increased human demand for water resources, together with the influence of climate change, has dramatically altered the water cycle and has degraded water conditions which support vegetation in many parts of the world, especially in arid and semiarid areas where the water cycle balance and vegetation ecology is fragile (Froend and Sommer, 2010). Vegetation canopy can be directly observed from space by remote sensing technology. The information of vegetation growth and distributions can be characterized from Normalized Difference Vegetation Index (NDVI) over time and across spatial scales (Seaquist et al., 2003; Gebremichael and Barros, 2006; Groeneveld and Baugh, 2007). The NDVI derived from remotely sensed data can effectively reveal the status of vegetation growth on the ground. It utilizes the contrast between the strong reflection of ground vegetation in the near infrared wavelength and the strong absorption by chlorophyll in the red wavelength (Gao and Dennis, 2001). The NDVI values reflect the density and greenness of the vegetation distributions. A few studies have been done to investigate the relationships between NDVI and some separate climatic variables. The relationships were commonly interpreted by using correlation analysis, transformed linear regression, and multiple-linear regression (Box et al., 1989; Schultz and Halpert, 1995; Ichii et al., 2002; Ji and Peters, 2004; Anyamba and Turcker, 2005). A number of studies showed that precipitation and NDVI were highly correlated (Richard and Poccard, 1998; Wang et al., 2001; Sarkar and Kafatos, 2004) and the correlations were varied from region to region. The overall relation between NDVI and precipitation is log-linear in Eastern Africa (Davenport and Nicholson, 1993) and linear in semiarid regions of Northeastern Brazil. However, there is no significant correlation in the Amazon watershed (Santos and Negri, 1997). Some climate-land interaction models at large regional scales were also used to assess the environmental impacts (Mao et al., 2012; West et al., 2014).

Comparing to the precipitation, the influence of groundwater on vegetation growth is usually less obvious and more difficult to be detected, especially during wet periods with a plentiful precipitation. Groundwater mainly gets recharged from precipitations and has great impact on the spatial and temporal distributions of soil moisture which further affects vegetation growth on the ground (Naumburg et al., 2005; Dai et al., 2014). NDVI has also been used to investigate the relationship between groundwater depth and vegetation at regional scales (Doble et al., 2006; Jin et al., 2007; Jin et al., 2014). Laboratory experiments and field investigations were conducted to investigate the effects of groundwater level drawdown on the performance of vegetation species (Kotowski et al., 2001; Froend and Sommer, 2010; Chen et al., 2014). Some approaches, such as the statistical model of curve-fitting regression procedure, were employed to assess the influence of groundwater on vegetation growth (Stromberg et al., 1996). Various mechanistic models, such as eco-hydrological models (Chui et al., 2011), vegetation competition models coupled with saturated-unsaturated hydrological models (Brolsma et al., 2010a, 2010b, Condon and Maxwell 2014, Tillman et al., 2012), EDYS models (Childress et al., 2002), WAVES models (Zhang et al., 1996), and IWSV model (Cheng et al., 2011) were developed to examine those relationships.

This study uses an arid and semiarid area, the Xiliao River Plain of northern China, as an example to develop an integrated framework to assess the impact of precipitation and groundwater on vegetation growth. The study area is located in Eastern Inner Mongolia of China with an area of 55, 378 km$^2$ (Figure 1). Forest land, grass land and cultivated land are the three major types of lands in this region. The major vegetation types include *Stipa Baicalensis*, *Grandis* and *Aneurolepidium Chinense*, *Agropyron Cristatum*, *Ulmus Pumila* and *Artemisia Halodendron*. The average annual precipitation in the 2000s varies from 178 mm in the western Naiman desert to 368 mm in the eastern region, while the average annual evaporation is 1900 mm for the same period. About 70% of the annual precipitation falls between June and August each year. Average annual temperature is between 5℃ and 6℃. The surface water system mainly includes the Xiliao river, Jiaolai river, Laoha river, Xilamulun river and Xinkai river. While there have been relevant studies in land cover change (Brogaard and Prieler, 1998), gross primary production (GPP) (Brogaard et al., 2005) and biomass production (Runnström, 2003), a quantitative investigation of the integrated impact of precipitation and groundwater on vegetation growth conditions is needed in this area. A better understanding of the water conditions which support vegetation growth will be helpful for a better management of the scarce water resources in this area.

This study examines the correlations among NDVI, precipitation and groundwater simultaneously by using their time-series data from 1981 to 2010. Numerical simulations of the interactions among precipitation, groundwater and plant water uptake are conducted to quantitatively evaluate the water conditions at the vadose zone of this area. The results provide insights which would help us to develop a strategy for regional water resources management and to mitigate the impact of the changing climate and the increased groundwater depth on vegetation growth.

## 2. Methodology

### 2.1 Integrated framework

Interactions between precipitation and groundwater and their contributions to vegetation growth are very complex. An integrated framework is developed to assess the relationships among vegetation, precipitation and groundwater depth, which systematically combines remote sensing technology with plant water uptake modeling in the vadose zone and the regional historical precipitation and groundwater measurements from 1981 to 2010 in the study area. The vegetation growth is quantitatively evaluated with the remote sensing data (by the NDVI) and the simulated plant water uptake rates. The correlations among precipitation, groundwater depth and the NDVI are investigated by using Pearson correlation equations. Numerical simulations of plant water uptake would further quantitatively evaluate the water sources (precipitation and groundwater) which support the vegetation growth. The flowchart of the integrated framework is showed in Figure 2.

### 2.2 NDVI data

Two NDVI datasets for the study area are downloaded from Web sites of Earth Resources Observation System Data Center (EROS) and Pathfinder Land Dataset (PAL). The first one is the level-3 NDVI product of Advanced Very High Resolution Radiometer (AVHRR) on National Oceanic and Atmospheric Administration satellites (NOAA). It covers the period from 1981 to 2000 at a ten-day interval and has a resolution of eight kilometers. The second one is the level-3 NDVI product of Moderate Resolution Imaging Spectro-radiometer (MODIS) named MOD13A3. It covers the period from 2001 to 2010 at a monthly interval and has a resolution of one kilometer. The MODIS data can be used directly, whereas NOAA data need to be processed for converting digital number (DN) to NDVI before application (Equation 1).

$$NDVI = 0.008 * (DN-128) \qquad (1)$$

The NOAA ten-day-interval NDVI data are composited into monthly NDVI data by using the maximum value composite (MVC) technique which can retain the highest NDVI value for each pixel and minimize cloud contamination as well as off-nadir viewing effects (Habib et al., 2009). Correlation analyses of NDVI with precipitation and groundwater are implemented within the period from 1981 to 2000 and the period from 2001 to 2010 separately in order to eliminate the potential calibration bias from different sensor systems (NOAA and MODIS).

The study area is one of the major grain-growing areas in northern China. About 20% of the area has been cultivated with well-developed irrigation facilities. Those cultivated areas have to be excluded from correlation analysis since irrigation activities will distort the natural relationship between precipitation, groundwater and vegetation. Landsat Thematic Mapper (TM) images are used to delineate the cultivated lands. Seven TM images are needed to cover the entire area. A maximum likelihood supervised classification is performed in ENVI (The Environment for Visualizing Images System) to map the cultivated land. The classification accuracy is about 94% (Zhao and Zhu, 2012). The cultivated lands are predominantly distributed along the river courses. Those cultivated areas were then excluded from correlation analysis.

## 2.3 Precipitation and groundwater data

The monthly precipitation data are collected from six meteorological stations within the study area. The groundwater depth data are collected from 14 long-term monitoring wells (Figure 3). Here the groundwater depth is defined as the distance from the ground surface to groundwater table. This analysis is focused on six analysis areas which are defined by the circles centred at six meteorological stations with a radius of 20 km. We focus on the time variety of the groundwater depth. From downstream to upstream, the groundwater spatial variability is small. The groundwater depth is interpolated into a grid with a resolution of 1 km by using the ordinary Kriging method with a spherical semivariogram model. The groundwater depth for each circled analysis area is then extracted from the grid by averaging all cell values within the analysis area. The NDVI values are also averaged in these analysis areas.

## 2.4 Correlation analysis of NDVI-precipitation and NDVI-groundwater depth

The Pearson's correlation coefficient is also called Pearson Product Moment Correlation (or PPMC), which is a measure of the strength and direction of the linear relationship between two parameters that is defined as the sample covariance of the variables divided by the product of their sample standard deviations, or

$$r = \frac{Cov(x, y)}{\sigma_x \sigma_y} = \frac{n\sum xy - \sum x \sum y}{\sqrt{\left[n\sum x^2 - (\sum x)^2\right]\left[n\sum y^2 - (\sum y)^2\right]}} \tag{2}$$

where, $Cov$ is the covariance, $\sigma$ is the standard deviation of the measured data $x$ or $y$. The Pearson correlation coefficient is calculated between the maximum NDVI of each circled analysis area and the accumulated precipitations of the maximum NDVI month and the prior two months. The correlation is also calculated between the maximum NDVI and groundwater depth at the same month in order to assess the relationship between groundwater depth and water condition of vegetation.

## 2.5 Numerical simulations of plant water uptake

The variably-saturated vadose zone in this area mainly consists of alluvial-proluvial fan sediments, including black calcium soil, silt sand, medium-fine sand, and gravel. From upstream to downstream of the Xiliao River, the aquifer strata changes from a single thick layer to multiple thin layers and the total thickness of aquifer layers becomes thinner. The groundwater heads of this region are reducing because of the increasing groundwater exploitation and the reduction of the annual precipitation in recent years. A numerical model for simulating plant water uptake is established by using a geological cross section near the observation well K1 in Kulun. The model has a width of 5 m, a depth of 3 m and a thickness of 1 m. The corresponding precipitation, evaporation and transpiration data (from 1981 to 2010) are assigned at the top atmosphere boundary. The bottom boundary is a variable water head boundary, in which the boundary water head data were obtained from the long-term measurements of the well K1. We assume that the water flow in the variably saturated porous media is three-dimensional isothermal Darcian flow and the flow equation is given by a modified form of the Richards' equation (Dai and Samper, 2004; Dai et al., 2008; Šimůnek et al., 2011):

$$\nabla(K_r K \nabla h) + w = (\phi \frac{\partial S_w}{\partial \psi} + S_w S_s)\frac{\partial \psi}{\partial t}, \tag{3}$$

where $h$ is hydraulic head which is the sum of pressure head $\psi$ and elevation Z

$$h = \psi + Z \tag{4}$$

Hydraulic conductivity $K$ is the product of relative conductivity $K_r$ and saturated conductivity $K_s$. $S_w$ is water saturation degree defined as the ratio between volumetric water content $\theta$ and porosity $\phi$, or $S_w = \theta/\phi$. Water saturation is related to pressure head through retention curve

$$S_w(\psi) = S_r + (1-S_r)[1+(-\alpha\psi)^n]^{-m} \tag{5}$$

where $S_r$ is the residual water saturation and $m$, $n$ and $\alpha(1/L^{-1})$ are Van Genuchten parameters usually estimated by fitting this function to observation data (van Genuchten, 1980). $w$ is a source or sink term, which include the plant (or root) water uptake, or the volume of water removed from specific volume of soil in a specific time due to plant water uptake. Feddes model (1974) and van Genuchten approach (1987) are used to calculate the root water uptake. Equation (3) is highly nonlinear because both hydraulic conductivity and saturation degree are functions of pressure head. The finite element numerical method combined with Newton-Raphson iteration scheme is used to solve the nonlinear Equation (3) (Šimůnek and Hopmans 2009; Šimůnek et al., 2008; 2011). The numerical model computes the plant water uptake, infiltration rate, and flow rate at water table, which will be used to analyze the intrinsic water cycle and the water budget for vegetation growth in the vadose zone.

## 3. Results and discussion

### 3.1 General pattern of NDVI, precipitation and groundwater depth

The NDVI value in the study area generally rises from west to east. This distribution corresponds to the regional precipitation pattern, which typically reflects the close relationship between vegetation and precipitation (Figure 4). From 1981 to 2000 the regional maximum NDVI has a mild increasing trend as shown in Figure 5, which might reflect an improved ground vegetation cover of this region. The precipitations in the 2000s, when the region experienced a dry period, are markedly lower than those in the 1980s and 1990s. The reduction of precipitations in the 2000s do not dramatically bring down the NDVIs (Figure 5) in the whole region. However, the lower precipitations in the 2000s apparently stop the upward trend of maximum NDVI.

Groundwater depths of the six analysis areas have a mild increase during 1981 to 2000. The increase has been sped up since early 2000s. The faster decline of groundwater table since early 2000s (Figure 6) corresponds to the lower precipitation during this period. A lower precipitation reduces the water supplement to the groundwater and encouraged more usage of groundwater for irrigation. In addition, an expanded population in the region has certainly put more pressure on the water resources. About 63% of the groundwater depth measurements are within 2 meters in the 1980s. In the 1990s the dominant groundwater depth values are from 2 to 3 meters while only 21.7% of the groundwater depth values are less than 2 meters. In the 2000s, this percentage is dropped to about 6%.

## 3.2 Relationship between NDVI and precipitation

The seasonal variation patterns of NDVI and precipitation in the six analysis areas are generally similar. The NDVI values starts rising in May and typically reached to the maximum value in August, then gradually declines afterward. Most rain falls in summer (June, July and August). The three-month accumulated precipitations have declined in most of analysis areas since the 1980s, while the NDVI values have slightly increased in all analysis, mostly occurs before the 2000s. A moderate increase of NDVI over a long period is a relatively common case in a well cultivated region because of the ongoing improvement of ground vegetation coverage. The dry period during the 2000s only slightly brings down regional NDVIs. A downward trend line of three-month accumulated precipitation is mainly driven by the lower precipitation during the 2000s.

The correlations between maximum NDVIs and the accumulated precipitations are shown in Table 1. The correlation coefficients are relatively weak, especially in Kulun and Keerqin with the coefficients of about 0.2 in 2000s and 0.3 from 1981 to 2000, which suggests that the precipitation is not a dominant factor and that the groundwater may have played an active role in those areas to maintain an effective water condition for vegetation growth. However the role played by precipitation may has been enhanced in the dry period of the 2000s when the four analysis areas had higher correlations between maximum NDVI and the three-month accumulated precipitation than those in the 1980s and 1990s (Table 1). For example, the coefficient in Kailu changed from 0.27 to 0.7, while the value in Kezuozhong changing from 0.45 to 0.61. The ground vegetation is more responsive to precipitation events under a depressive dry condition.

## 3.3 Relationship between NDVI and groundwater depth

Some of the vegetation species are very sensitive to groundwater depth and others may be intensive. The relationship between NDVI and groundwater depth can reflect the average behaviour of all vegetation types to groundwater depth (Jin et al., 2014). There is a strong correlation between the maximum NDVI and groundwater depth in Keerqing and Kulun during the 2000s dry period, as shown in Table 1, while both areas have the weakest correlations (with a coefficient of about 0.2) between precipitation and NDVI during this period, which suggests that there is a close connection between the groundwater depth and vegetation growth in these two areas. Ground vegetations have the ability to adapt to the water conditions by changing its composition.The capillary ability of the groundwater vegetation root may be strengthened. The groundwater apparently plays an effective role to sustain a soil moisture condition for vegetation growth of these areas during this long dry period. In Kezuozhong, NDVI is negatively correlated to groundwater depth with a coefficient of -0.54 at the p-values of 95 %, which suggests there is also a connection between the groundwater condition and vegetation growth in this area. No significant correlation is found between groundwater depth and NDVI in the rest three analysis areas (Kailu, Kezuohou and Zhalute).

**3.4  Annual precipitation, groundwater depth and the maximum NDVI**

Annual precipitation, groundwater depth and the maximum NDVI are investigated in wet years (precipitation event occurs at less than 15th percentile of all precipitation event) and in dry years (precipitation event occurs at exceeding 85th percentile of all precipitation event) in order to make a further assessment of the complex relationships among these three elements.

Precipitation is usually the principal supplier of soil moisture and is therefore the major factor to affect the water supply for vegetation growth. The maximum NDVI values in wet years are greater than that in dry years in Kailu (Figure 7a) and Zhalute areas. More precipitation could raise the NDVI under the similar groundwater depth condition. For example, the NDVI values in 1982 and 1998 were 0.43 and 0.57 with the similar groundwater depth of about 2.1m in Kailu. This phenomenon also occurs in Kezuohou and Kezuozhong. In addition, precipitation is the main source for groundwater recharge, higher precipitations raises the groundwater tables in Kailu, Keerqin and Kezuohou. The groundwater depths in these regions during wet years are shallower than that in dry years.

Groundwater can supply a certain amount of water to sustain an efficient soil moisture condition for vegetation growth and support vegetation to survive through a long dry period. In that case lower annual precipitations sometimes do not bring down the NDVI values, such as in Keerqin and Kulun areas (Figure 7b). Groundwater has played an active role in these two areas particularly during the prolonged drought in the 2000s. Both areas have a considerably low NDVI in year 2009 as these two areas suffered a significant reduction of groundwater heads after a prolonged drought.

### 3.5 Plant water uptake modeling

An obvious correlation between the NDVI and groundwater depth are observed in Keerqing and Kulun during the 2000s dry period. While Keerqing is a highest cultivated area in the study area, the influence of the irrigation may not be totally removed due to the resolution of the TM image. Kulun is chosen as a typical place to build the numerical model for quantitatively assessing the plant water uptake and the recharge water sources (i.e., precipitation and groundwater). The geological cross section and the finite element mesh for numerical simulations are showed in Figure 8. The flow model parameters listed in Table 2 are collected from local infiltration experiments and literatures (Dai et al., 2008; Šimůnek et al., 2011).

Figure 9 shows the numerical simulation results. The computed actual plant water uptake (Figure 9a) is from two sources: actual infiltration from precipitation (Figure 9b) and recharge from groundwater (Figure 9c). Note that the infiltration rates are the actual water (the precipitation subtracts the surface flow and evaporation) infiltrated into the vadose zone. It is obvious that the infiltration water from precipitation is directly uptaken by plant roots. If there is more infiltration than plant can uptake, it may recharge groundwater (Figure 9b).

Figure 9c shows the water budget at the water table from 1981 to 2010. The positive flow rates are that the infiltration water recharges groundwater while the negative flow rates are that groundwater recharges to the vadose zone to sustain the soil moisture that is needed for vegetation growth. In particular, during the low precipitation periods (from 2001 to 2010, Figure 9c) the groundwater can recharge the vadose zone to preserve efficient soil moisture to sustain the growth of vegetation, and to support ground vegetation to survive through a drought. A correlation analysis result indicates that the groundwater has an increased contribution to the plant water uptake during that drought period.

On the other hand, the contribution from groundwater to the vegetation growth would be reduced when precipitations are plentiful or groundwater table is fallen. The supplemented water from groundwater storages would be reduced to a negligible level when the groundwater table has fallen below a certain level, where the plant root cannot reach. Therefore a fallen groundwater table may pose a threat to the survival of vegetation during a prolonged drought. Maintaining proper groundwater depth is vital to vegetation growth, in particular to vegetation survival during a long dry period in arid and semiarid regions, which is a more crucial issue when there is an increasingly volatile precipitation pattern under a changing climate.

## 4. Conclusions

An integrated investigation of remotely sensed data, precipitation and groundwater data can help to achieve a better understanding of the relationships among precipitation, groundwater and ground vegetation in arid and semiarid regions. This study has developed an integrated framework to assess the relationships among precipitation, groundwater depth and NDVI by using the field data collected from 1981 to 2010 in the study area. The spatial distribution of NDVI generally matches the regional spatial pattern of precipitation, which indicates that precipitation is a major factor to govern the distribution of natural vegetation in the region.

Obvious correlations between groundwater depths and NDVIs exist in parts of the study area during a prolonged drought in the 2000s, which reflects a close connection between the groundwater depth and vegetation growth in these areas. The numerical simulation results of the plant water uptake quantitatively demonstrate that precipitation infiltration and groundwater alternatively provide water for vegetation growth in different seasons. The infiltration water from precipitation is directly uptaken by plant roots. If there is more infiltration than plant can uptake, it may recharge groundwater. During low precipitation periods the groundwater can help to preserve efficient soil moisture to sustain the growth of vegetation, and to support ground vegetation to survive through a drought. The contribution from groundwater to the vegetation growth would be reduced as the groundwater depth increased. Maintaining proper groundwater depth is vital to vegetation growth in arid and semiarid regions, which has been a more crucial issue with an increasingly volatile precipitation pattern under a changing climate. The derived results provide information for the local government and policy makers to improve the management of water resources in the study area.

A more detailed interpolation of the spatial precipitation surface from more precipitation gauges will deliver a more accurate precipitation value at each analyzing location, while a denser groundwater monitoring well network will provide more accurate groundwater measurements for the study area. That will improve the accuracy and reliability of the correlation analysis between NDVI, precipitation and groundwater. In addition, different vegetation types may response to regional water conditions differently. For instance, while bush land and grassland typically have a rapid and strong response to a precipitation event, permanent vegetation such as forests will have a much delayed and sometime unobvious response. Note that water is the key factor in vegetation growth, but is not the only factor. Other environmental and climatic conditions (such as temperature and surface elevation) do have influences on vegetation growth. Implementing a more comprehensive analysis with multiple factors simultaneously is sometime practically not feasible, in particular, is not the best option to understand the impact from the specific factors. Further study will include more environmental factors in the integrated assessment.

**Acknowledgements**

This work was supported by National Natural Science (No.41201420, 41130744) , Beijing Nova Program (No.Z111106054511097) and Beijing Young Talent Plan. The authors are thankful to Xinyin Cui of the Songliao Water Resource Committee for providing the field data.

**List of Tables**

Table 1: Correlation coefficients between three-month accumulated precipitation, groundwater depth and maximum NDVI

| Relationship | | Kailu | Keerqin | Kezuohou | Kezuozhong | Kulun | Zhalute |
|---|---|---|---|---|---|---|---|
| Precipitation-NDVI | 1981-2000 | 0.27 | 0.34 | 0.36 | 0.45 | 0.32 | 0.31 |
| | 2001-2010 | 0.70 | 0.21 | 0.43 | 0.61 | 0.21 | 0.51 |
| Groundwater depth-NDVI | 2001-2010 | -0.33 | -0.60 | -0.23 | -0.54 | -0.64 | -0.22 |

Table 2: Numerical modeling parameters in the cross-section (modified from

| Cross-section layers | Thickness(m) | Residual water content | Saturated water content | $\alpha$ (1/m) | $n$ | $Ks$(m/d) |
|---|---|---|---|---|---|---|
| Black calcium soil | 0.5 | 0.1 | 0.43 | 0.5 | 1.5 | 0.3 |
| Silt sand | 1 | 0.08 | 0.41 | 0.8 | 1.6 | 0.5 |
| Fine sand | 1 | 0.06 | 0.38 | 1.2 | 1.8 | 1 |
| Sandy clay | 0.5 | 0.09 | 0.4 | 1.1 | 1.4 | 0.2 |

**List of figure captions**

Figure 1:   Location of the study area overlapped with Landsat TM image using RGB color system. The black line is county boundary, the blue line represents river and the white dot means county. The green color reflects plant cover.

Figure 2:   Flowchart of the integrated framework

Figure 3:   Locations of six analysis areas

Figure 4:   Distribution patterns of the maximum NDVI in 1980s (a), 1990s (b) and 2000s (c)

Figure 5:   The precipitation, maximum and minimum NDVI during the vegetation growing season in the whole region

Figure 6:   The fluctuation of groundwater depth in six analysis areas from1981 to 2010

Figure 7:   Annual precipitation, groundwater depth and the maximum NDVI in dry and wet years (a: Kailu, b: Kulun)

Figure 8:   A cross-section for the plant water uptake modeling (a) and the numerical mesh (b)

Figure 9:   The numerical simulation results for (a) plant water uptake, (b) actual infiltration rate, and (c) flow rate at groundwater table

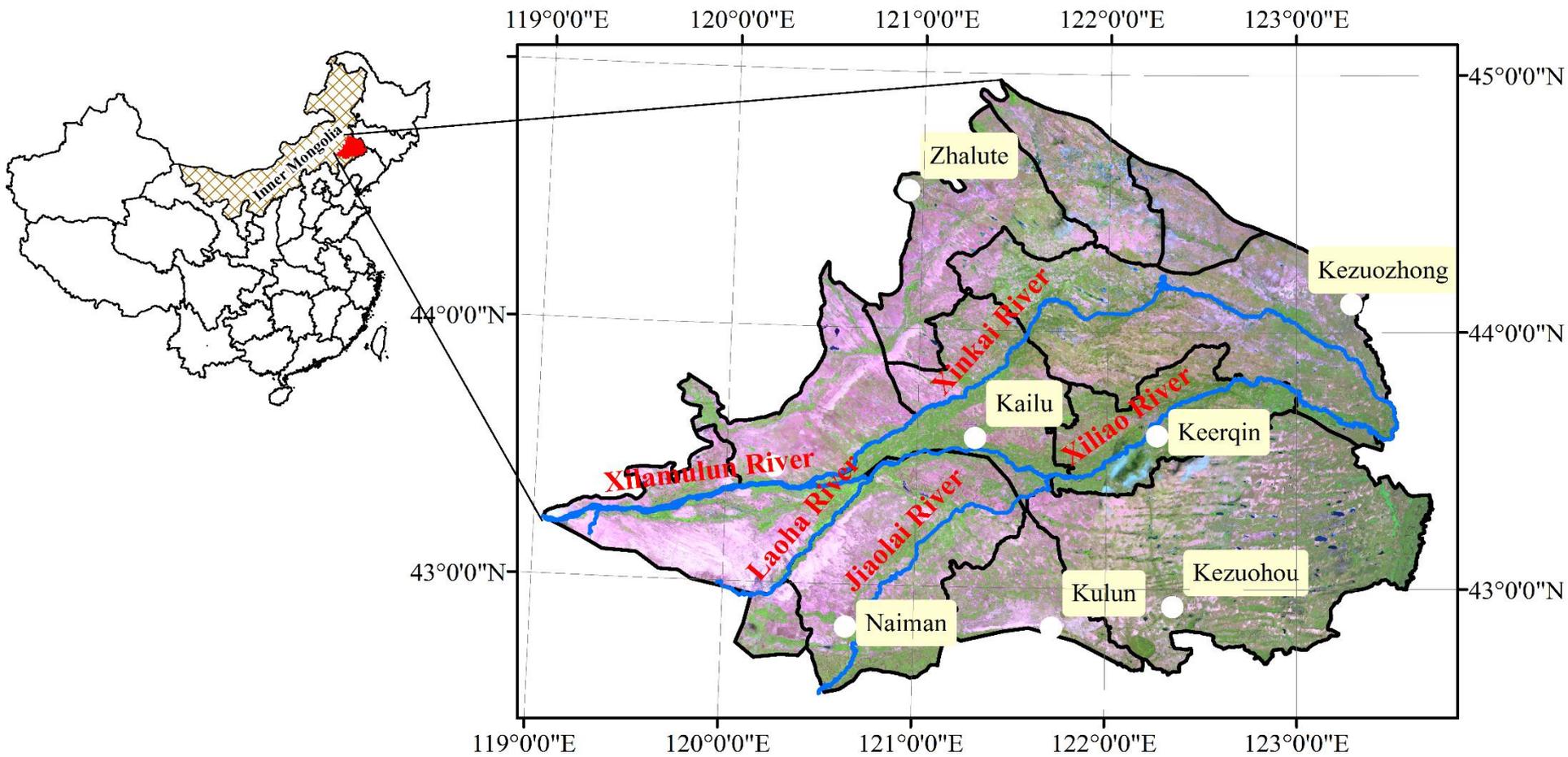

Figure 1

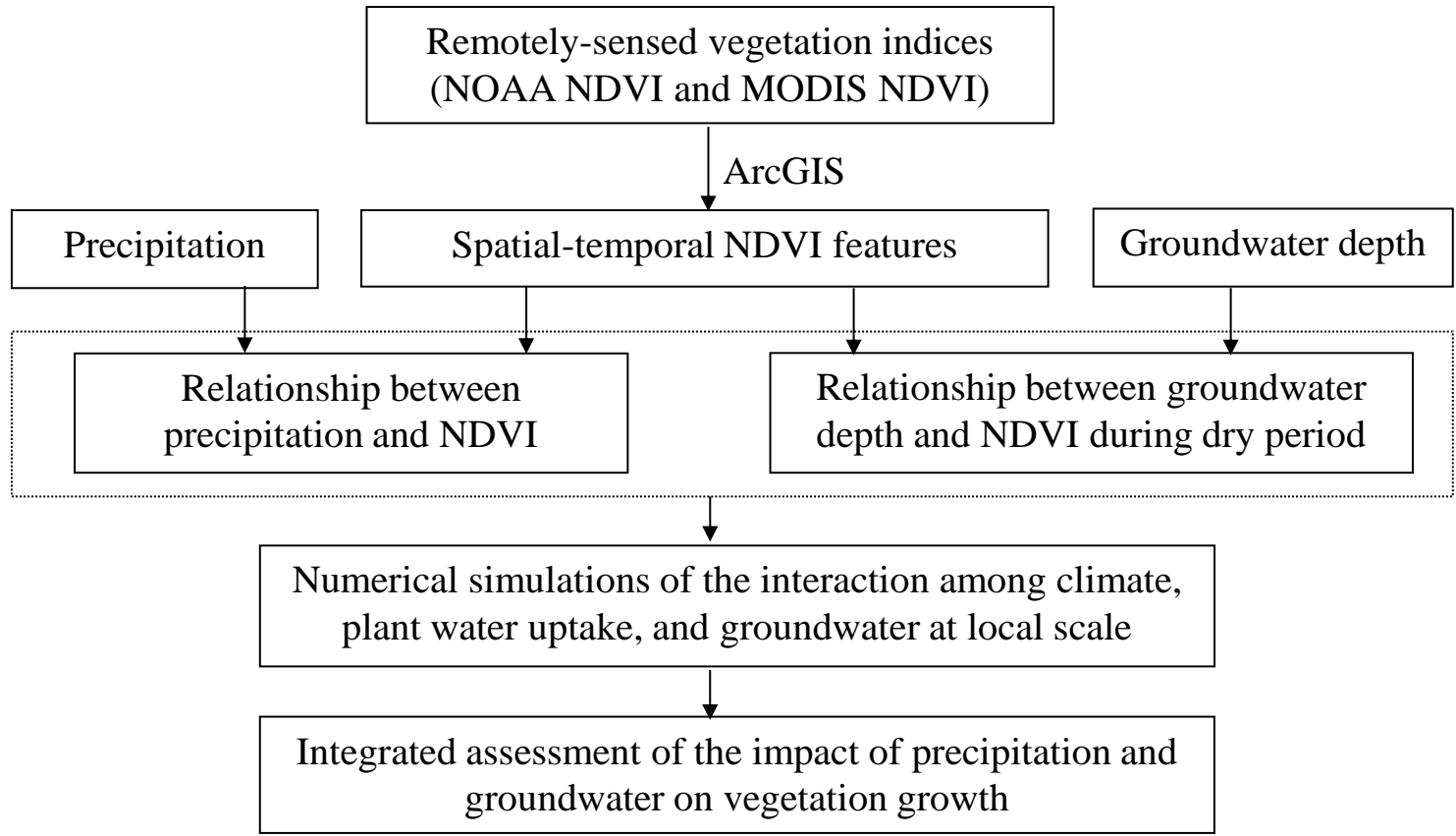

Figure 2

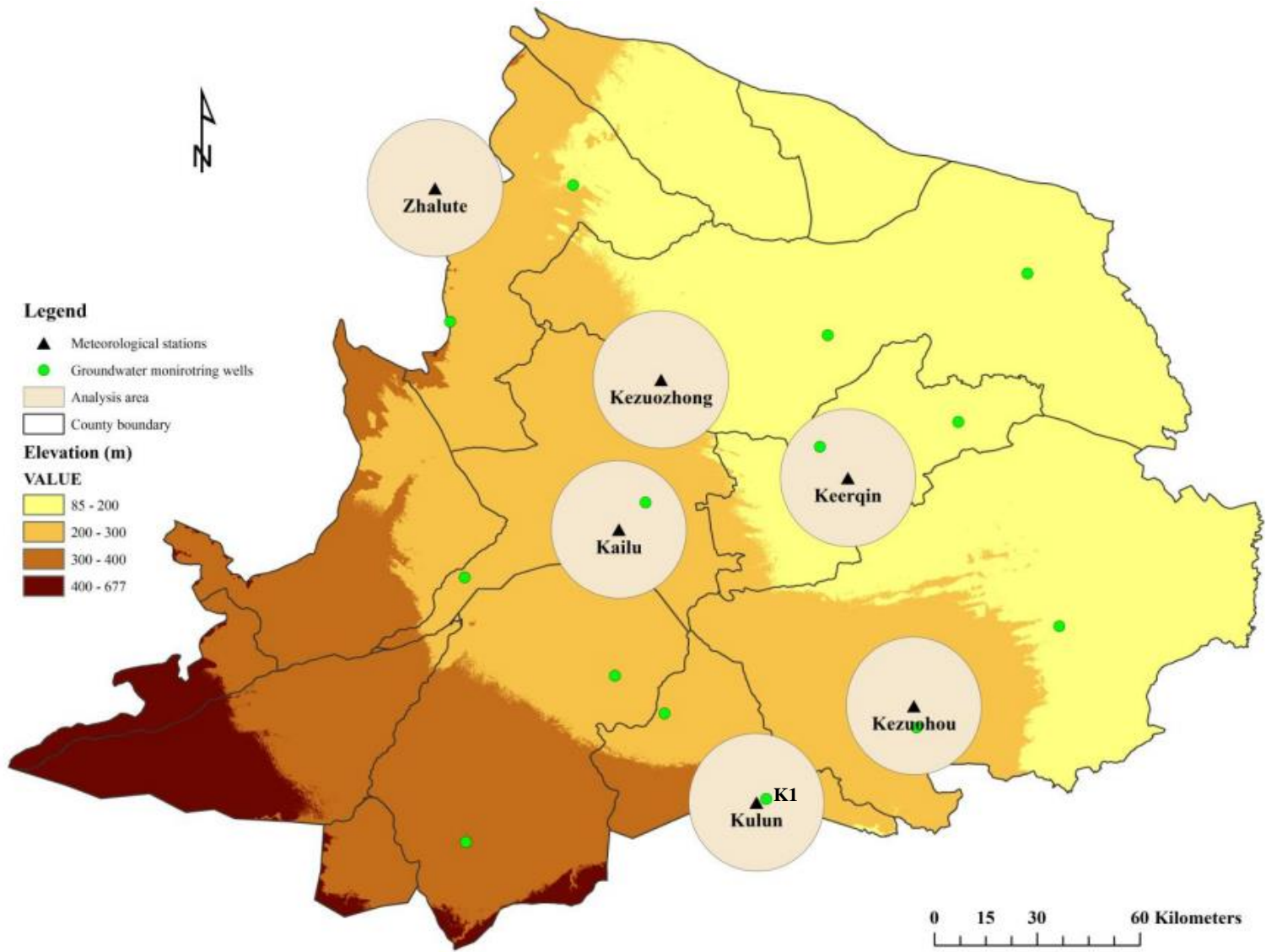

Figure 3

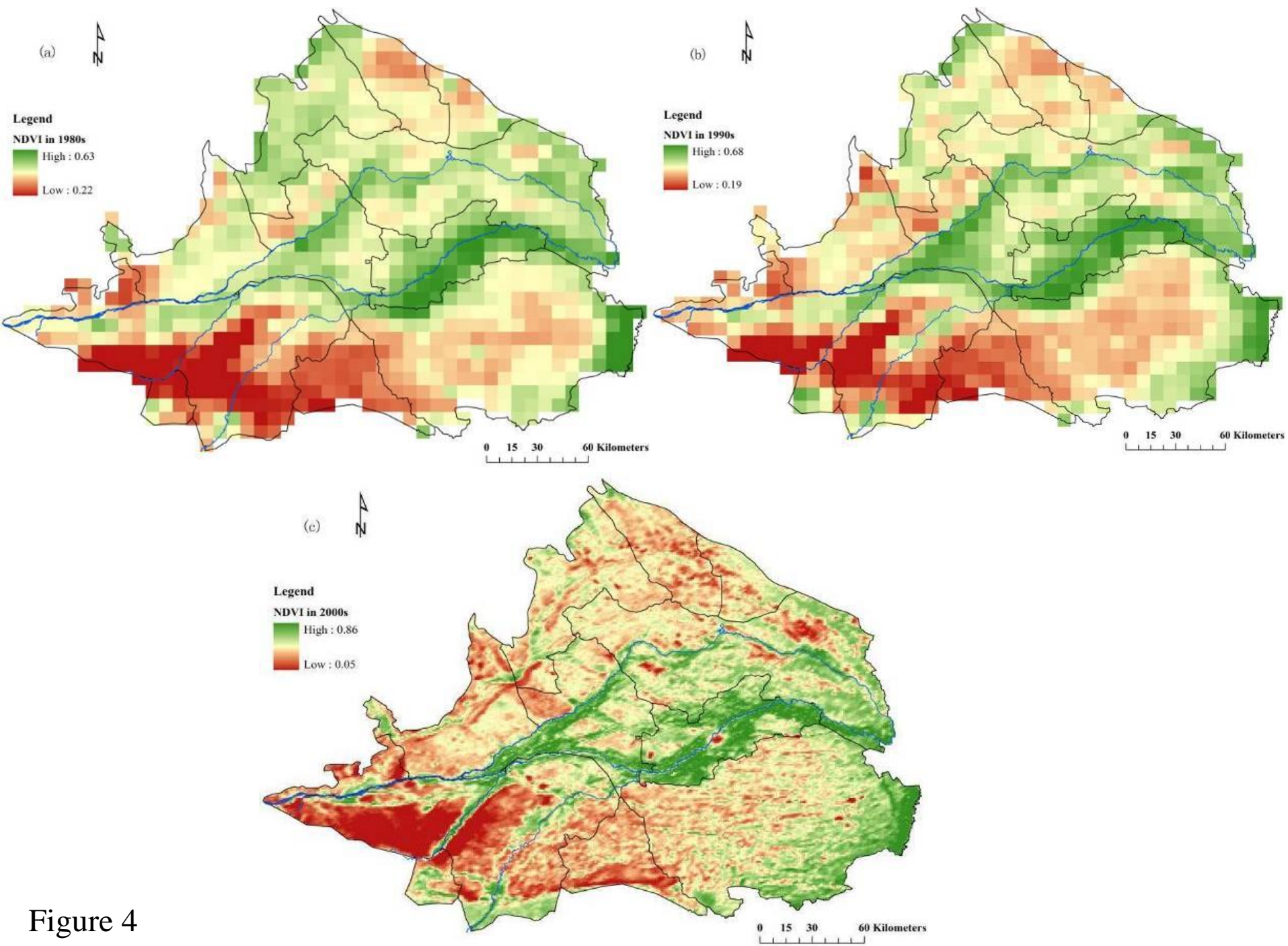

Figure 4

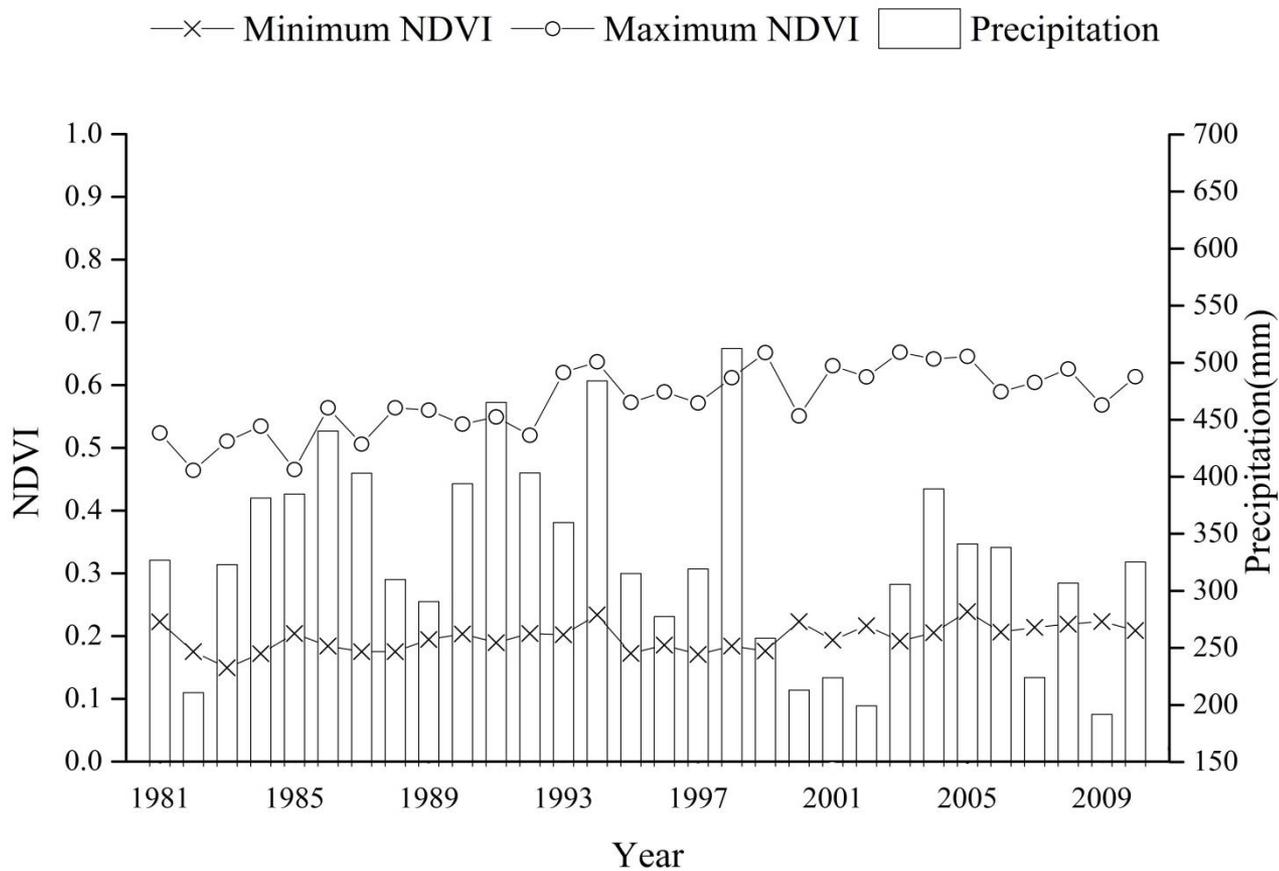

Figure 5

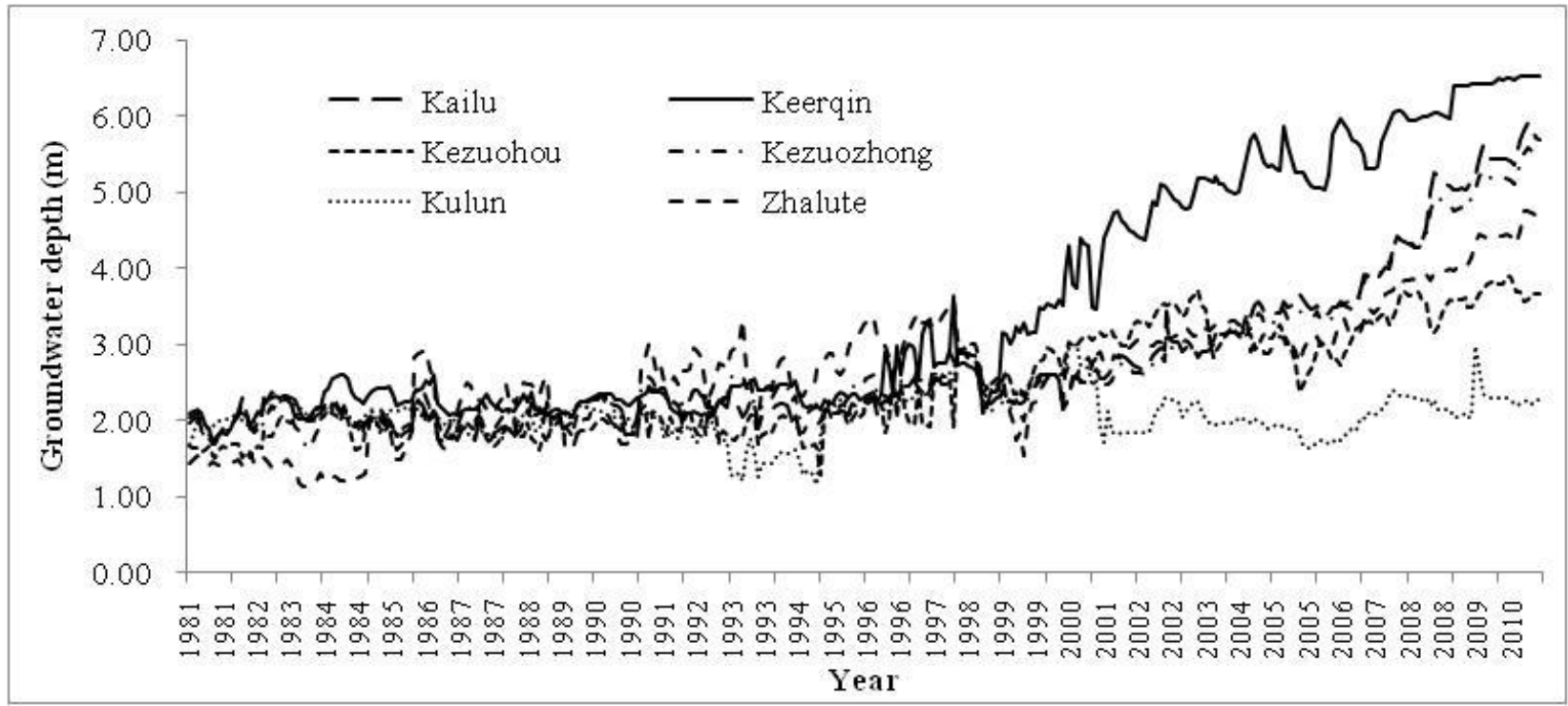

Figure 6

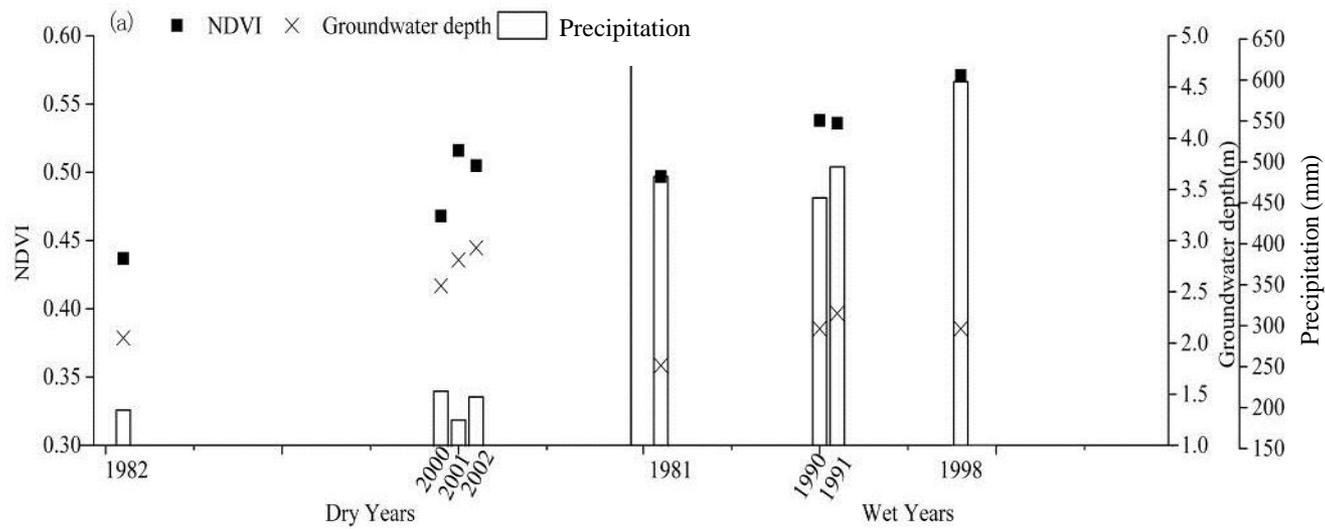

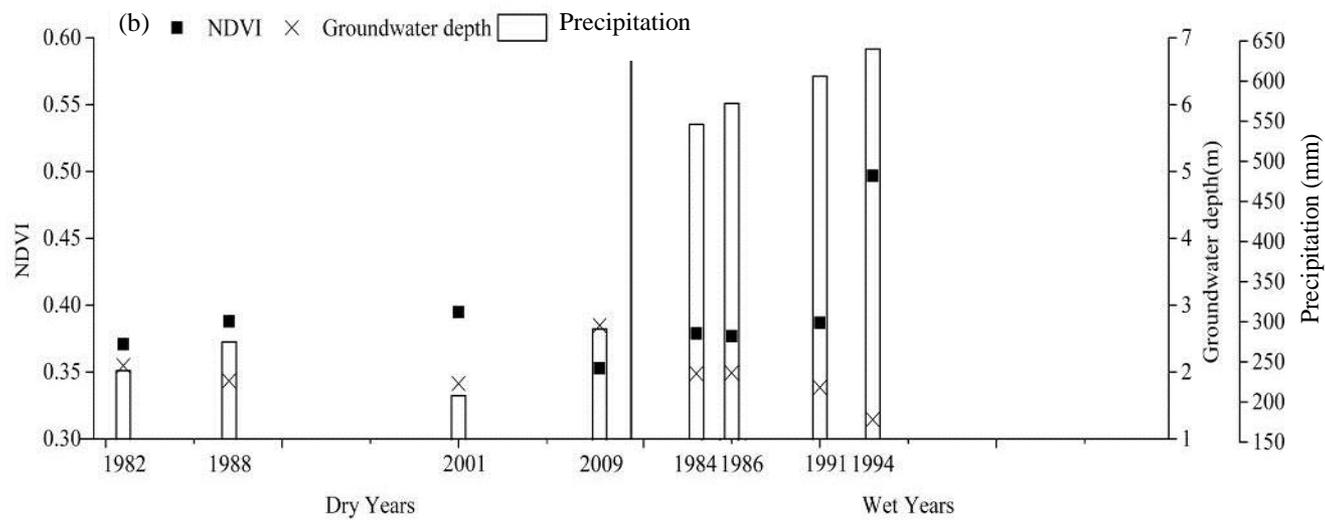

Figure 7

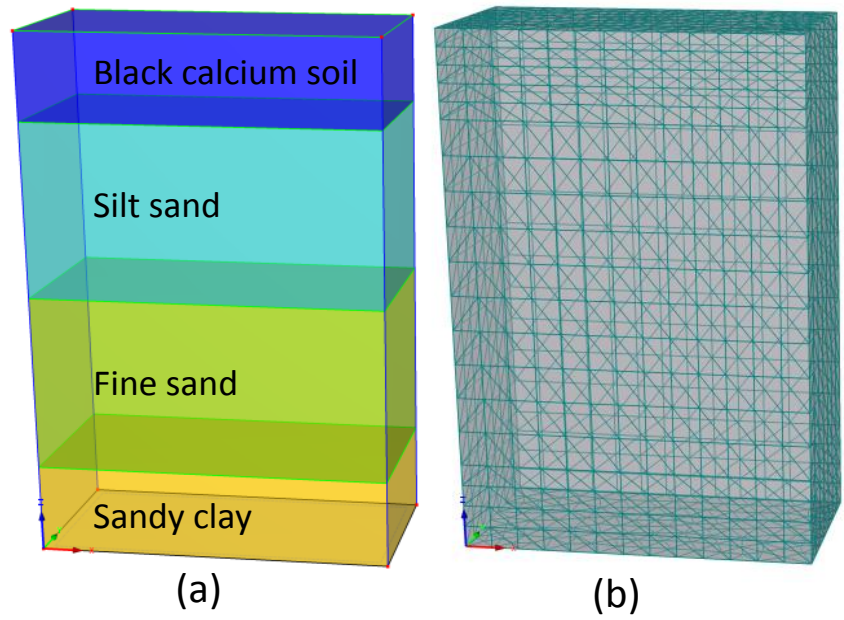

Figure 8

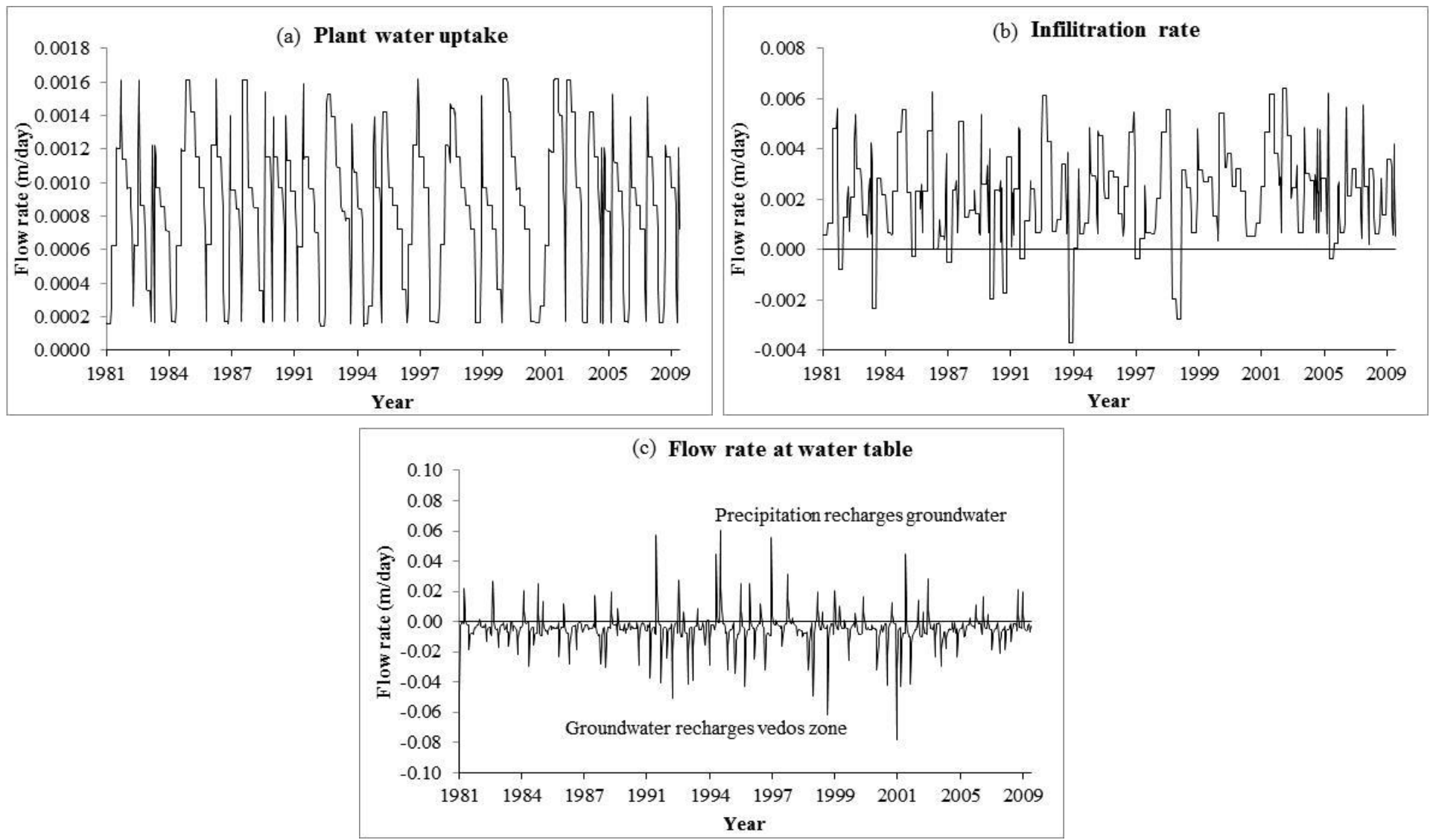

Figure 9